\documentclass{svjour2}                    % onecolumn
\smartqed  % flush right qed marks, e.g. at end of proof
\usepackage{graphicx}
\begin{document}

\title{The  transition towards immortality: non-linear autocatalytic growth of citations to scientific papers
%\thanks{Grants or other notes
%about the article that should go on the front page should be
%placed here. General acknowledgments should be placed at the end of the article.}
}
%\subtitle{Do you have a subtitle?\\ If so, write it here}

%\titlerunning{Short form of title}        % if too long for running head

\author{Michael Golosovsky         \and
        Sorin Solomon %etc.
}

%\authorrunning{Short form of author list} % if too long for running head

\institute{
             The Racah Institute of Physics, the Hebrew University of Jerusalem, Jerusalem 91904, Israel \\
              Tel.: +972-2-6586551\\
              Fax: +972-2-5617805\\
              \email{michael.golosovsky@mail.huji.ac.il}           %  \\
%             \emph{Present address:} of F. Author  %  if needed
           \and
%           S. Author \at
 %             second address
}

\date{Received: date / Accepted: date}
% The correct dates will be entered by the editor

\maketitle

\begin{abstract}
We discuss microscopic mechanisms of complex network growth, with the special emphasis of how these mechanisms can be evaluated from the measurements on real networks. As an example we consider the network of citations to scientific papers. Contrary to common belief that its growth is determined by the \emph{linear} preferential attachment, our  microscopic measurements show that it is driven by the \emph{nonlinear autocatalytic growth}. This invalidates the scale-free hypothesis for the citation network. The nonlinearity is responsible for a dramatic dynamical phase transition: while the citation lifetime of majority of papers is 6-10 years, the highly-cited papers have practically infinite lifetime.
\keywords{power-law distribution \and citations \and preferential attachment \and complex networks \and autocatalytic growth}
\PACS{01.75.+m, 02.50.Ga, 89.75.Fb, 89.75.Da }
% \subclass{MSC code1 \and MSC code2 \and more}
\end{abstract}

\section{General introduction}
A lot of  empirical evidence for the power-law degree distribution in  natural networks has been amassed during last decade.  This led to the conjecture that these networks are scale-free. It is widely believed that the growth of the scale-free networks is driven by  the cumulative advantage mechanism \cite{Price} which is commonly known as the preferential attachment  \cite{Barabasi,Newman_SIAM}. This mechanism assumes that  $\Delta k$, the  number of  links  acquired by a node during a short time interval $\Delta t$ is determined by the number of already acquired links $k$, 
\begin{equation}
\Delta k= A(k+k_{0})
\label{rate_naive}
\end{equation}
Here, $k$ is the node degree, $k_{0}$ is the  "initial attractivity" and  $A$ is the attachment rate (aging function) which is time-dependent. Equation \ref{rate_naive} yields the power-law degree distribution,  $P(k)\propto 1/k^{\gamma}$, which is generally considered as a fingerprint of a scale-free network.  The linear preferential attachment (Eq. \ref{rate_naive}) is believed to be one of the most important microscopic mechanisms that generates the scale-free complex networks which are so ubiquitous in nature.

This statement is often reversed  and the power-law degree distribution in a growing network is considered as an evidence for the linear preferential attachment. The parameter $k_{0}$ is  estimated from the exponent of the degree distribution \cite{Barabasi}:
\begin{equation}
\gamma=2+\frac{k_{0}}{m}.
\label{gamma}
\end{equation}
where $m$ is the mean degree.  This approach meets several difficulties. First of all it yields unrealistically high $k_{0}\approx m$. Second  and most important- the validity of the power-law approximation for degree distribution in complex networks has been contested. Indeed, since the node degree is a discrete and non-negative number, the scale-free power-law function cannot  provide a good fit for the nodes with small degree. At best, it can fit only the fat tail of the  distribution. However, several recent studies showed that the degree distribution in complex networks can deviate from the power-law dependence even in the fat tail \cite{Redner,Amaral,Fortunato,Petersen,Golosovsky_EPL}. 

Krapivsky and Redner \cite{KRL} showed that the deviation necessarily occurs if the  attachment  kernel is nonlinear,
\begin{equation}
\Delta k= A(k+k_{0})^{\alpha}
\label{rate_nonlinear}
\end{equation}
In particular, for sublinear attachment kernel, $\alpha<1$  the network is characterized by the stretched exponential degree distribution; while for the superlinear  kernel, $\alpha>1$,  the network organizes into a "winner takes all" configuration \cite{KRL,Dorogovtsev2002}. While for linear attachment kernel the network achieves stationary degree distribution, for the nonlinear case  the degree distribution is nonstationary. In the sequel we  call  the dynamics governed by Eq. \ref{rate_nonlinear} as the "nonlinear autocatalytic growth" \cite{Ofer99,Blank2000} and reserve the term "preferential attachment" for  Eq. \ref{rate_naive} that generates the power-law degree distribution. 

 To what extent the growth mechanism of real networks deviates (Eq. \ref{rate_naive}) is  an important question. Recent experimental studies \cite{Redner,Jeong,Eom2008,Eom2011,Valverde,Csardi,Wang,Tomassini,Newman,Capocci,Bingol,Eisenberg,Sheridan} that measured microscopic growth of complex networks, came up with the conclusion that the attachment exponent $\alpha$ is close to unity, in such a way that the growth mechanism is nearly linear (see Table I) and thus follows Eq.\ref{rate_naive}. However, to which extent  $\alpha$ deviates from  unity remained  an open question until now. The above studies could hardly measure this deviation  due to  time-dependence of $\alpha$, finite precision limited by the size of their databases and, most important - due to uncertainty arising from the use of different methodologies. In particular, Ref. \cite{Sheridan} applied four different methods to measure attachment kernel in the  network of the US patent-to-patent citations and found different exponents ranging from 1.12 to 1.38.

\begin{table}
\caption{\label{tab:table1} Testing preferential attachment in real networks}
\label{tab:1}
\begin{tabular}{llll}
\hline\noalign{\smallskip}
Network&Ref.&Attachment &Method\\
&&exponent $\alpha$&\\
\noalign{\smallskip}\hline\noalign{\smallskip}

%Citations to scientific papers &  & & \\
Citations & Jeong et al. \cite{Jeong} &0.95&cumulation\\
of scientific papers& Eom $\&$Fortunato \cite {Eom2011} &1&cumulation\\
& Redner \cite{Redner} &0.9-1.05&running average\\
& Wang et al. \cite{Wang} &1&running average\\
& Golosovsky $\&$Solomon \cite{Golosovsky_PRL} &1-1.25& histogram\\
&&(\emph{grows with time})&\\
%US patent citations & &&\\
 & &&\\
Citations & Csardi et al. \cite{Csardi} &1.2&histogram\\
of US patents& Sheridan et al. \cite{Sheridan} &1.23-1.27&Metropolis-Hastings\\
& Valverde et al. \cite{Valverde} &1-1.25&cumulation\\
&&(\emph{grows with time})&\\
  & &&\\
%Scientific collaboration  networks & &&\\
Scientific &Jeong et al. \cite{Jeong} &0.8&cumulation\\
collaboration & Tomassini $\&$Luthi \cite{Tomassini} &0.76&cumulation\\
& Newman \cite{Newman} &1&histogram\\
 & &&\\
Movie actors & Jeong et al. \cite{Jeong} &1&cumulation\\
 & Eom et al. \cite{Eom2008} &1&cumulation\\
  & &&\\
Wikipedia& Capocci et al. \cite{Capocci} &0.76&histogram\\

%Internet  & &&\\
Google& Jeong et al. \cite{Jeong} &1.05&cumulation\\
Internet& Eom et al. \cite{Eom2008} &1&cumulation\\
Internet Dictionary & Herdagdelen et al. \cite {Bingol} &1&histogram\\
Protein networks & Eisenberg $\&$Levanon \cite{Eisenberg} &1&cumulation\\
\noalign{\smallskip}\hline
\end{tabular}
\end{table}

The goal of our present study is the high precision measurement of the microscopic growth rate of a complex network and the determination of the attachment exponent $\alpha$. Following the accepted practice  \cite{Redner,Jeong,Eom2011,Wang}, as an object of our research we chose one of the best-documented complex networks: citations to scientific papers. Here, the papers are nodes and  citations to these papers are links. We performed high-statistics and time-resolved study of the citation dynamics of a very large and homogeneous set of papers. In what follows we compare two methods of measuring the microscopic growth rate of this  network: averaging (histogram) and cumulation. We found that the former method is quite reliable and yields superlinear attachment kernel, $\alpha\approx 1.25$, while the latter method is prone to quantization errors. We came to conclusion that the microscopic growth mechanism of the citation network follows \emph{nonlinear} autocatalytic growth  (Eq. \ref{rate_nonlinear}) with $\alpha>1$.
%\cite{Newman,Redner,Wang,We_PRL} \cite{Jeong,Eom2008,Eom2011,Tomassini}

We elaborate on a  dramatic consequence of nonlinearity: if one considers a citations dynamics governed by the superlinear attachment kernel, one is led to conclusion that  this network contains a subset of the papers  that will be cited forever. Thus we witness a dynamical phase transition in which citation lifetime of a paper diverges to infinity. Our measurements provide experimental evidence for such runaway papers that have practically infinite  citation lifetime.

\section{Methodology}
\begin{figure*}
\includegraphics[width=0.75\textwidth]{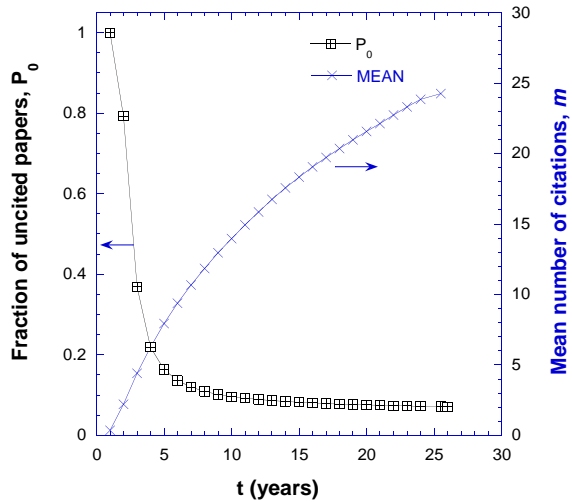}
\caption{Time dependence of the fraction of uncited papers $P_{0}$ and of the mean number of citations $m$ for 40195 Physics papers published in 1984. $t$ is the number of years after publication, whereas the publication year corresponds to $t=1$. The continuous lines are guide to the eye. While  the number of uncited papers saturates after 15 years, the mean number of citations does not saturate even after 25 years.
}
\label{fig:uncited-mean}
\end{figure*}

To assess the microscopic growth mechanism  of the citation  network we focussed on one discipline- Physics. We considered a cohort of papers  published in the same year $T_{publ}$ and measured the number of citations garnered by each paper in every subsequent year  $T_{cit}$. To this end we used  the Thomson-Reuters ISI Web of Science, chose 82 leading Physics journals, excluded review articles, comments, editorial, etc.,  and analyzed  citation history of  40,195  original research  papers  published in these journals in $T_{publ}=1984$. The cumulative citation distributions for this data set were demonstrated elsewhere \cite{Golosovsky_EPL}. Figure \ref{fig:uncited-mean} shows some aggregate characteristics of this set: the mean number  of citations  and the fraction of uncited papers.

\begin{figure*}
\includegraphics[width=0.8\textwidth]{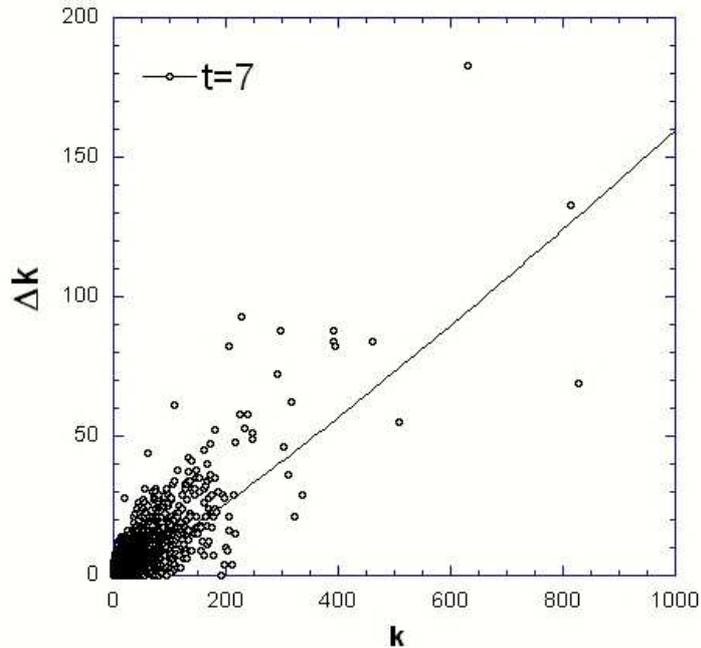}
\caption{The scatter plot of the number of additional  citations $\Delta k_{i,7}$, garnered by each paper $i$ during seventh year after publication. The horizontal axis shows $k_{i,6}$ -the  total number of citations  garnered by the same paper during six previous years. The solid line displays approximation by Eq. \ref{rate}  with $\alpha=1.13, k_{0}=1,A=0.065$.
}
\label{fig:spread}
\end{figure*}

In what follows we focus on two variables: (a) $k_{i,t}$ - the cumulative number of citations, i.e. the total number of  citations accumulated  by a paper  $i$ in the period between $T_{publ}$ and  $T_{cit}$;  and (b) $\Delta k_{i,t+\Delta t}$ - the number of additional citations gained by the same paper  in a short time window  between $T_{cit}$ and $T_{cit}+\Delta t$. Here,  $\Delta t=1$ year and  $t=T_{publ}-T_{cit}+1$ (if $T_{publ}=T_{cit}$ then $t=1$, in such a way that $k_{i,1}$ measures the number of citations during the year when the paper was published).  Figure \ref{fig:spread}  shows  $\Delta k_{i,t+1}$ versus $k_{i,t}$.  (Specifically,  $k_{i,6}$ is the number of citations  garnered by a paper $i$ from 1984 to 1989 while $\Delta k_{i,7}$ is the number of citations garnered by the same paper in 1990.) While the trend of increasing $\Delta k_{i,t+1}$  versus $k_{i,t}$ is clearly visible, the  fluctuations are so strong  that Fig. \ref{fig:spread} does not provide an obvious proof of the validity of Eq. \ref{rate_naive}.

This is not unexpected since the actual number of newly acquired citations is a stochastic variable. We define  $\lambda_{i}(t)=\overline{\Delta k_{i,t}}/\Delta t$  which is the average citation rate over the ensemble of the nodes with the same $k_{i,t}$. The autocatalytic growth model  actually claims that $\lambda_{i}=A(k_{i}+k_{0})^{\alpha}$, in such a way that
\begin{equation}
\Delta k_{i,t}=A(k_{i}+k_{0})^{\alpha}\Delta t +\sigma dW(t)
\label{rate}
\end{equation}
where $\sigma dW(t)$ is a random variable with zero mean and $\sigma^{2}$ variance (for brevity we replace thereon $t+1$ by $t$). In contrast to $\Delta k_{i,t}$ which is a discrete variable,  $\lambda_{i}(t)$ is a continuous one.  To verify whether the noisy data, such as those shown in Fig. \ref{fig:spread}, are generated by the growth law suggested by Eq. \ref{rate}, there have been developed two methods: averaging (histogram) and cumulation. The processing of our data using  these methods yielded conflicting results. In what follows we compare these two methods and  develop a control tool to check their internal consistency.

\section{Comparison between different methods to measure the microscopic growth law of citation network}
\subsection{\textbf{Histogram (Averaging) Method}}
To infer the microscopic growth law from the noisy data such as those shown in Fig. \ref{fig:spread},  one  bins the data, finds the mean  $\lambda=\overline{\Delta k_{i}}$ for each bin, and compares the resulting histogram to the prediction  of Eq. \ref{rate}. This approach was first  used by Newman \cite{Newman} to verify the linear preferential attachment hypothesis in real networks.  The Refs. \cite{Csardi,Capocci,Bingol} followed this approach as well, while  Refs. \cite{Redner,Wang} used a very similar  moving average procedure.
\begin{figure*}
\includegraphics[width=0.75\textwidth]{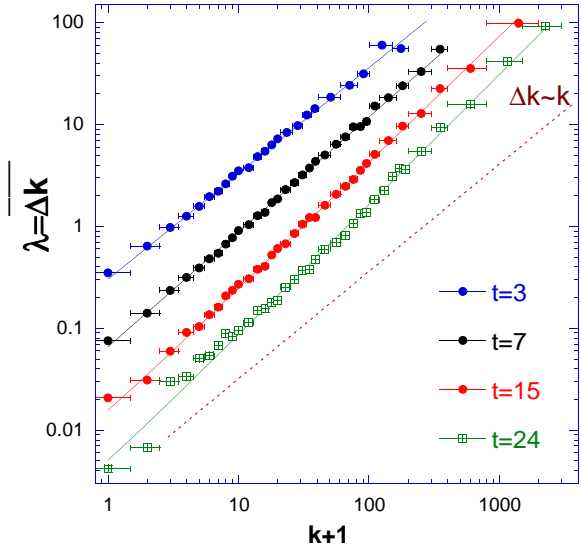}
\caption{Mean number of additional citations, $\lambda(k)=\overline{\Delta k_{i}}$, as a function of the  number of previous citations  $k(t)$; $t$ is the number of years after publication. The additional citations are counted in the time window of $\Delta t=1$ year. To include uncited papers ($k=0$)  the horizontal axis displays $k+1$ instead of $k$. Each set of points corresponds to a certain citing year. The straight dashed line shows linear approximation  $\overline{\Delta k}\propto (k+k_{0})$ where $k_{0}=1$. The data deviate upwards from this linear dependence, especially at $t=15-24$. The continuous lines show better, superlinear fits, $\overline{\Delta k}=A(k+k_{0})^{\alpha}$ where $A,\alpha$ and $k_{0}$ are fitting parameters.  The superlinear dependences fit  the highly-cited papers ($k>100$) and uncited papers ($k=0$) as well. 
}
\label{fig:all_mean}
\end{figure*}

To process our data in such a way we chose a certain citing year $T_{cit}$, grouped all papers into $\sim$ 40 logarithmically-spaced bins, each bin containing the papers with close $k$, found the  mean number of next year citations $\lambda$ for each bin and plotted it versus $k$. Figure  \ref{fig:all_mean} shows such $\lambda(k)$ dependences. In particular, the black circles  indicate the results of  the averaging procedure applied to the data of the Fig. \ref{fig:spread}. The $\lambda(k)$ dependences are fairly well fitted by  Eq. \ref{rate}.

\begin{figure*}
\includegraphics[width=0.75\textwidth]{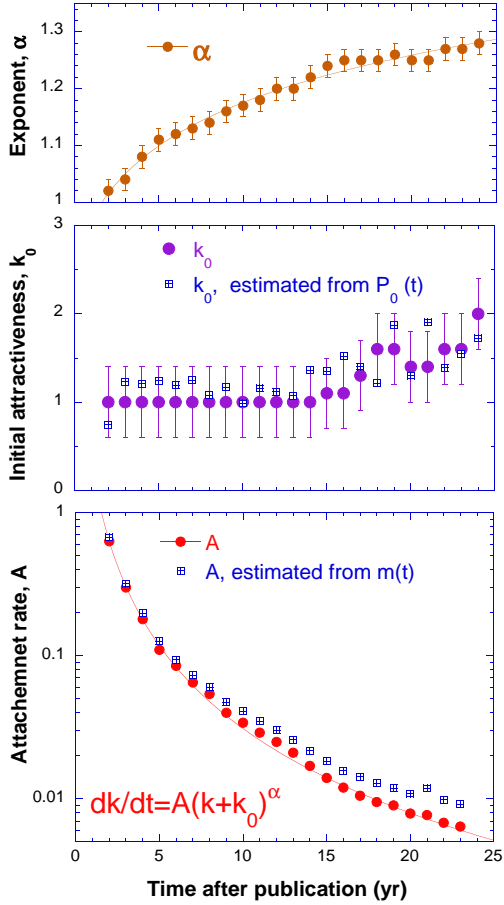}
\caption{ Time dependence of the parameters  of Eq. \ref{rate}. (a) Exponent $\alpha$. (b) Initial attractivity $k_{0}$. (c) Rate constant $A$. The continuous  line  in (a) is a guide to the eye while in (c) it shows empirical approximation $A=3.3/(t+0.3)^{2}$ where $t$ is the number of years after publication.  The  blue squares in (b) and (c) show the estimates of $k_{0}$ and $A$ based on Eqs. \ref{A},\ref{k_0} and  Fig. \ref{fig:uncited-mean}, correspondingly. The consistency of $k_{0}$ and $A$  obtained by two methods [circles vs squares] validates the \emph{superlinear } preferential attachment, $\alpha>1$. 
}
\label{fig:mean}
\end{figure*}

Figure \ref{fig:mean} shows time dependence of  the fitting parameters $\alpha,k_{0},A$. The exponent $\alpha$ gradually increases with time from $\alpha=1$ to $\alpha=1.28$, indicating linear attachment kernel for "young" papers and superlinear attachment kernel for "old" papers. The initial attractiveness $k_{0}\approx 1.1$ is almost time-independent and is surprisingly close to ad hoc assumption of de Solla Price \cite{Price}. The time dependence of the attachment rate $A$ follows the empirical power-law dependence, $A=3.3/(t+0.3)^2$. A similar power-law dependence can be inferred from the US patent citation data of Ref. \cite{Csardi}.

\subsection{\textbf{Cumulation Method}}
Jeong, Neda, and Barabasi \cite{Jeong} were the first to measure the  growth rate of evolving networks used the cumulation method. This method quickly became the most popular tool to assess the preferential attachment in real networks \cite{Eom2008,Eom2011,Valverde,Tomassini,Eisenberg,Sheridan}. It consists in calculation of the kernel $\kappa(k)=\int_{0}^{k}\Delta k' dk'$ where $k'$ is the total number of citations garnered by a paper by year $T_{cit}$ and $\Delta k'$ is the number of citations accrued by this paper during time window between $T_{cit}$ and $T_{cit}+\Delta t$ where $\Delta t$ is usually 1 year. The integration is performed over all papers that garnered $k'\leq k$ citations by year $T_{cit}$.  The key assumption behind this scheme is that the fluctuations in $\Delta k$ are averaged out and the resulting integral is the same as if Eq. \ref{rate} were integrated directly over $k$ at fixed $t$ i.e.,
\begin{equation}
\kappa(k)=\frac{A}{\alpha+1}\left[(k+k_{0})^{\alpha+1}-k_{0}^{\alpha+1}\right]
\label{integration}
\end{equation}

\begin{figure*}
\includegraphics[width=0.75\textwidth]{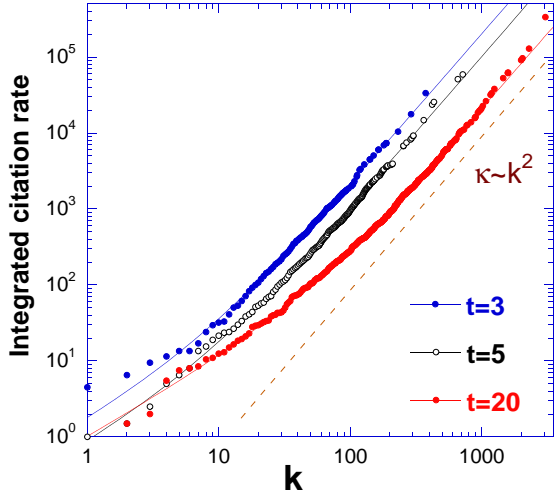}
\caption{ Integrated number of additional citations $\kappa(k)=\int_{0}^{k}\Delta k'dk'$, as a function of the  number of previous citations  $k$; $t$ is the number of years after publication.  We used the same raw data as those shown in Fig. \ref{fig:spread} and applied  trapezoidal numerical integration routine implemented by MATLAB.  The dashed line shows quadratic dependence $\kappa\propto k^2$ as expected for the linear preferential attachment. The data at high $k$ follow this dependence as if they were generated by the linear preferential attachment,  $\alpha\approx 1$. The continuous lines show fit given by Eq. \ref{integration} with $\alpha=1$ and $A$ and $k_{0}$ as fitting parameters.
}
\label{fig:integrated}
\end{figure*}

\begin{figure*}
\includegraphics[width=0.75\textwidth]{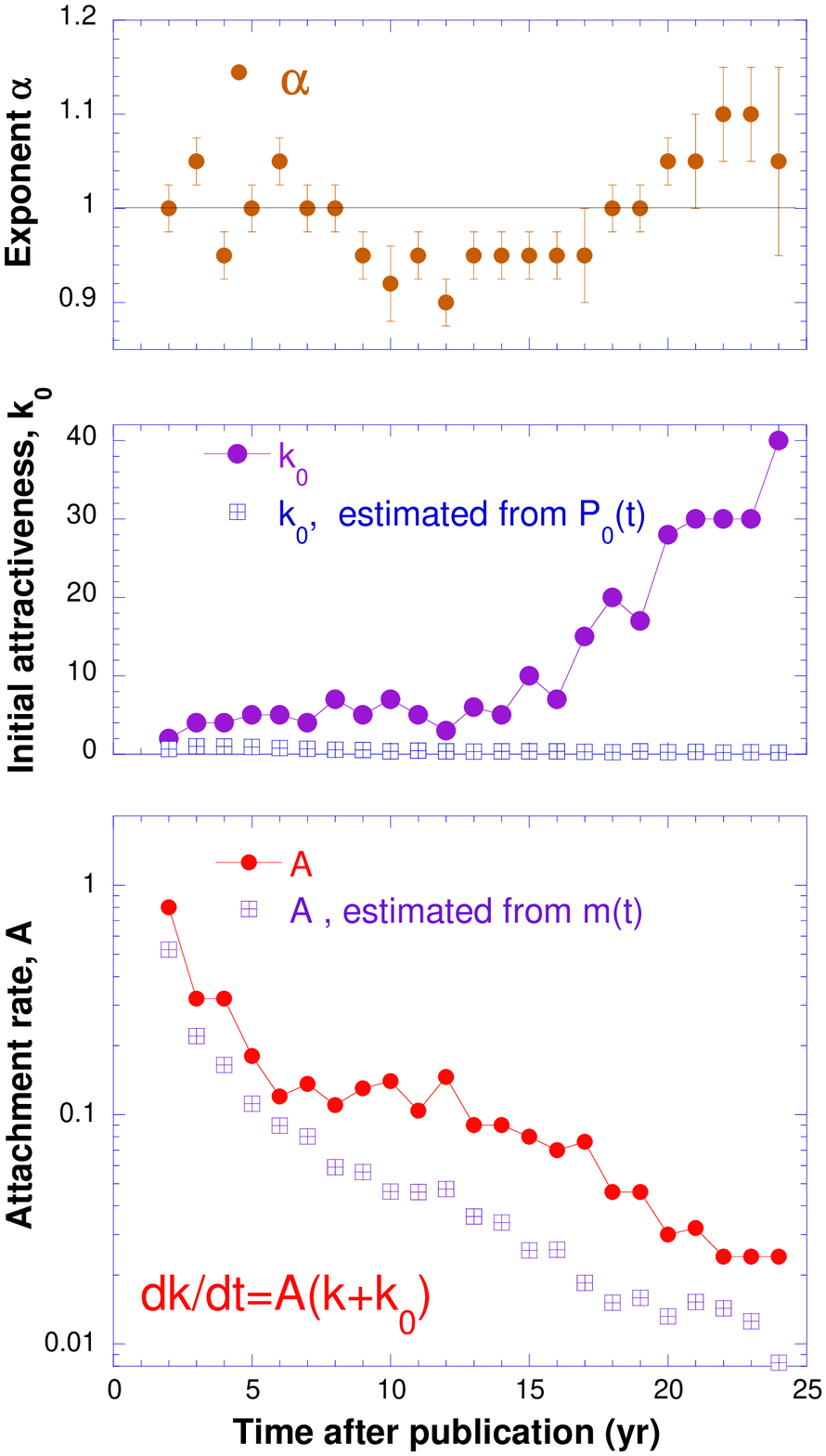}
\caption{ Time dependence of the parameters of  Eq. \ref{rate} as found using numerical integration  (Eq. \ref{integration}). (a) Exponent $\alpha$. (b) Initial attractivity $k_{0}$.  (c) The rate constant $A$.   The blue squares in (b) and (c) show, correspondingly, the estimates of $k_{0}$ and $A$ based on Eqs. \ref{A}, \ref{k_0} and the data of the Fig. \ref{fig:all_mean}.  The results  obtained by two methods [circles vs squares] strongly differ. This casts doubt on the validity of the cumulation method as applied to citations and especially on its  claim of the linear preferential attachment, $\alpha=1$.}
\label{fig:integration_three}
\end{figure*}

Figure \ref{fig:integrated} demonstrates application of this method  to our data.  The fluctuations have been dramatically reduced, as expected.  Equation \ref{integration} fits well the data for high $k$, while for low $k$ the fit is less satisfactory. Figure \ref{fig:integration_three} shows the fitting parameters. We found that the deviation of the exponent $\alpha$ from unity is within the experimental uncertainty, $\Delta \alpha=\pm0.05$. Therefore, to find  $A$ and $k_{0}$ we set $\alpha=1$ in our fitting procedure.
 
Figure \ref{fig:integration_three} shows  that the fitting parameters found in such a way are notably different from those found by the averaging method (Fig. \ref{fig:mean}). Most important - the exponent $\alpha$  is close to unity while that found from the averaging method is higher than unity. The initial attractivity $k_{0}$ is high and increases with time, while that found from the averaging method is close to unity and almost time-independent. The attachment rate $A$ exceeds that found from the averaging method, especially  at long times. The discrepancy between the two methods   calls for some control tool. In what follows we develop such tool and use it to decide which method: averaging or cumulation  is more reliable.

\subsection{\textbf{Control Tool}}
We consider here an additional tool to estimate the microscopic growth parameters of a growing network. We have developed this indirect method  to check the internal consistency of the  histogram and cumulation methods. This control method is based on two assumptions: (i) the microscopic growth law given by Eq. \ref{rate} is valid for all papers including uncited ones, and (ii) the exponent $\alpha$ is known. Then, the microscopic parameters  $A$ and $k_{0}$  may be estimated from the dynamics of the macroscopic parameters: the mean  number of citations $m$, and the fraction of uncited papers  $P_{0}$.

The mean, $m=\overline{k_{i}(t)}$, is the average number of citations garnered by a paper during the period between $T_{publ}$ and  $T_{cit}$, in such a way that the age is $t=T_{cit}-T_{publ}+1$. The averaging here is performed over all papers. Differentiation with respect to time yields the average number of additional citations garnered by a paper between $t$ and $t+\Delta t$, namely, $\frac{dm}{dt}\Delta t=\overline{\Delta k_{i}}$. We average Eq. \ref{rate} over all papers and for $\Delta t=1$  we find $\frac{dm}{dt}=A\overline{(k+k_{0})^{\alpha}} \approx A(m+k_{0})^{\alpha}$ (the last approximation holds  because $\alpha$ is close to unity). This yields the rate constant
\begin{equation}
A\approx\frac{dm}{dt}\frac{1}{(m+k_{0})^{\alpha}}
\label{A}
\end{equation}
For $k=0$ Eq. \ref{rate} reduces to $\lambda_{0}=Ak_{0}^{\alpha}$ where $\lambda_{0}$ is the average citation rate of  previously uncited papers. The latter can be recast through the fraction of uncited papers $P_{0}$ as follows, $\lambda_{0}\approx\frac{1}{P_{0}}\frac{dP_{0}}{dt}$  \cite{uncited}. Equation \ref{A} yields then
\begin{equation}
k_{0}\approx \left(\frac{1}{AP_{0}}\frac{d P_{0}}{dt}\right)^\frac{1}{\alpha}
\label{k_0}
\end{equation}
We solve  Eqs. \ref{A},\ref{k_0} for known $\alpha$ and find $A$ and $k_{0}$. If $\alpha$ is found properly the parameters $A$ and $k_{0}$ obtained by this control method shall be consistent with those found directly.

For the histogram (averaging) method, the $A$ and $k_{0}$ found from Eqs. \ref{A},\ref{k_0} are indeed consistent  with those found by the direct procedure (Fig. \ref{fig:mean}). The difference in $k_{0}$  is within the measurement uncertainty, while a small difference in $A$ can be traced to the Jensen's inequality, $\overline{x^{\alpha}}>\overline{x}^{\alpha}$ for $\alpha>1$.
However,  for the cumulation procedure,  the $A$ and $k_{0}$ found from Eqs. \ref{A},\ref{k_0} are clearly inconsistent  with those found directly (Fig. \ref{fig:integrated}): $A$  is substantially lower and  $k_{0}$ is also much smaller than those found directly. This inconsistency calls for a deeper consideration of the validity of the cumulation method as applied to citations.

 While the cumulation method works well  for noisy continuous data with  Gaussian fluctuations,  its applicability  to citations is problematic. Since the additional citations $\Delta k$ are discrete and non-negative, their fluctuations around the mean are non-symmetrical and their magnitude is on the order of the mean  (see Fig. \ref{fig:spread}). It appears that the standard numerical integration procedure as implemented, for example, in MATLAB does not work well for  discrete, wildly fluctuating data that have non-symmetrical distribution around the mean. The straightforward application of the numerical integration procedure (cumulation) for quantifying  dynamics of growing networks is thus ineffective. This method  shall be specially tailored for the discrete variables with the non-Gaussian and strongly skewed fluctuation spectrum.

We conclude that the microscopic parameters of the growing citation network obtained by the averaging (histogram) method are most reliable.

%
%\widetilde{k_{0}}

\section{The effect of the exponential growth of publications on citation dynamics}
Most theoretical studies consider networks that grow linearly in time. In fact, they define a "network time" in such a way that new nodes are added to network at constant rate. It should be noted that citation networks grow exponentially with time. In what follows we define the network time for the citation network  and recast our results in terms of the network time.

\begin{figure*}
\includegraphics[width=0.6\textwidth]{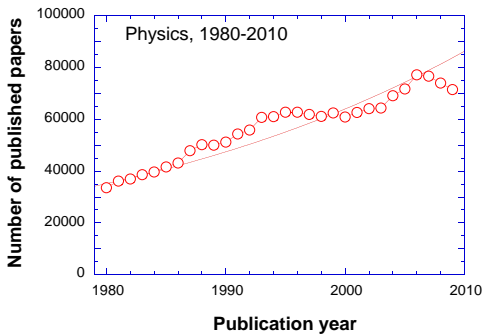}
\caption{ The annual number of published Physics research papers (not including conference proceedings). The solid line shows exponential approximation corresponding to 1.3$\%$ annual growth.}
\label{fig:annual_publ}
\end{figure*}

Figure \ref{fig:annual_publ} shows the annual growth of the number of original research papers and reviews in 82 leading Physical journals as covered by the ISI Thomson-Reuters Web of Science (excluding editorials, conference proceedings, etc.). The network growth is close to exponential, hence the difference between the physical time and network time is essential.

We define the "network year" $\Delta t^{*}$ in such a way that the number of papers published during this time interval is equal to the number of papers published in 1984. (This is approximately equivalent to time rescaling, $t^{*}=\frac{e^{0.032t}-1}{0.032}$ where $t$ is the physical time.) To determine microscopic parameters of the citation dynamics we  use Eq. \ref{rate} where $\Delta t=1$ year is replaced by $\Delta t^{*}$.  The data shown in Fig. \ref{fig:all_mean} are unaffected by this transformation although now they refer to  network time which is different from physical time. The parameters $\alpha$ and $k_{0}$ remain the same. The new rate constant  $A^{*}$   is determined from the slope of the dependences shown in Fig. \ref{fig:all_mean} divided by $\Delta t^{*}$ instead of $\Delta t$. 

\begin{figure*}
\includegraphics[width=0.6\textwidth]{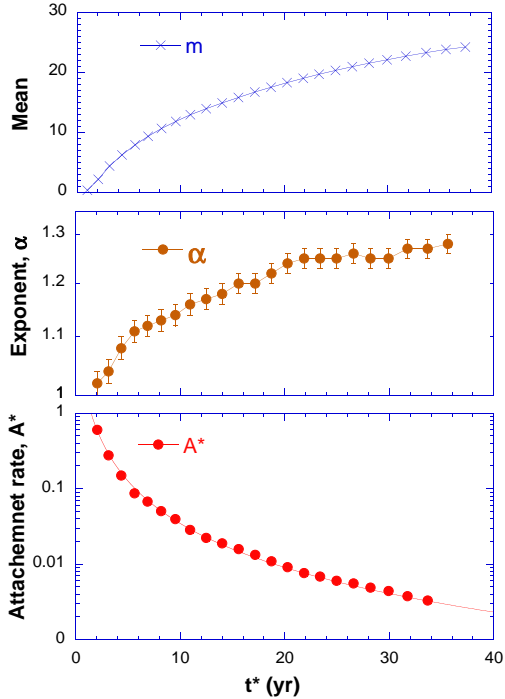}
\caption{ Microscopic parameters of the citation dynamics versus network time $t^{*}$. The latter is defined in such a way that the citation network grows linearly with $t^{*}$. (a) $m$, the mean number of citations. (b) $\alpha$, exponent of the attachment kernel.  (c) $A^{*}$, the rate constant. The continuous line  shows empirical approximation $A^{*}=3.8/(t^{*}+0.5)^{2}$. }
\label{fig:network}
\end{figure*}

Figure \ref{fig:network} shows some citation parameters in this time frame. The saturation exhibited by $m$ and $\alpha$  is clearly visible. This should be compared to Figs.\ref{fig:uncited-mean}, \ref{fig:mean}a  where the growth of $m$ and $\alpha$ is more close to logarithmical. The attachment rate $A^{*}(t^{*})$ turns out to be almost the same as the $A(t)$ dependence. This coincidence may be occasional.

We conclude that the growth of the citation network follows  Eq. \ref{rate} with time-independent $k_{0}\approx 1$, time-dependent exponent $\alpha$  which gradually increases from  1 to 1.28, and  the rate constant that decays with time as $\sim 1/t^{2}$.  This conclusion  remains the same if we replace the physical time by the network time. By the way, the studies of US patent-to-patent citations by the histogram method  yielded similar attachment exponent $\alpha\sim 1.2-1.27$ \cite{Valverde,Csardi,Sheridan}. All this indicates that  the citation networks experience the autocatalytic growth  with the \emph{superlinear} attachment kernel. Although the nonlinearity is weak, it leads to far-reaching consequences which we analyze below.

\section{Divergence of the citation lifetime - dynamical phase transition towards immortality}
In what follows we analyze the consequences of the \emph{nonlinear} growth mechanism of the citation network.  In particular, we demonstrate that the nonlinearity is responsible for the  enormous spread of citation lifetimes of scientific papers. To show this we consider citation dynamics of individual papers and distinguish between  the initial period of $t_{0}=$2-3 years when a paper makes an immediate impact  and the subsequent period when the citation dynamics of this paper to some extent is built on its initial success.   The  time dependence of the total citation count of a paper during this later period can be crudely estimated by integrating  Eq. \ref{rate} with respect to time as if $k$ were a continuous variable. In what follows we focus only on the papers with $k>>1$ (moderately- and highly-cited papers), in such a way that the term $k_{0}$ in Eq. \ref{rate} can be neglected. We also neglect for a moment the stochastic component of $k$. Assuming time-independent exponent $\alpha$ we integrate Eq. \ref{rate}   and find
%(The setting of $t_{0}$ to 3 years is somewhat arbitrary. The further analysis can be extended to any value of $t_{0}$).
\begin{equation}
k(t)=\frac{\widetilde{k_{0}}}{\left(1-\delta \widetilde{k_{0}}^{\delta}\int_{t_{0}}^{t}Adt\right)^{1/\delta}}.
\label{nonlinear}
\end{equation}
Here, $\widetilde{k_{0}}=k(t_{0})$ stands for the number of citations garnered by a paper during initial period  $t_{0}$ (it shouldn't be mixed with $k_{0}$ that appears in Eq. \ref{rate}), and $\delta=\alpha-1$. For $\delta<<1$ Eq. \ref{nonlinear}   reduces  to a more transparent form
\begin{equation}
k(t)=\widetilde{k_{0}}e^{\widetilde{k_{0}}^{\delta}\int_{t_{0}}^{t}Adt} \label{exp1}
\end{equation}
Consider two cases. 
\begin{enumerate}
    \item  $\delta=0 $  -linear growth. The $k(t)$ dependence  can be factorized, $k(t)=\widetilde{k_{0}}\dot e^{\int_{t_{0}}^{t}Adt}$, where  the $\widetilde{k_{0}}$ sets the scale and the factor $e^{\int_{t_{0}}^{t}Adt}$ sets the time dependence of the citation count. Citation dynamics of all papers should follow  the universal dependence $e^{\int_{t_{0}}^{t}Adt}$ which is independent of $\widetilde{k_{0}}$.
    \item $\delta\neq 0 $  -nonlinear growth mechanism. In this case Eq. \ref{exp1} can not be factorized and both the scale and the time dependence of the total citation count  depend on $\widetilde{k_{0}}$.
\end{enumerate}    

\begin{figure*}
\includegraphics[width=0.75\textwidth]{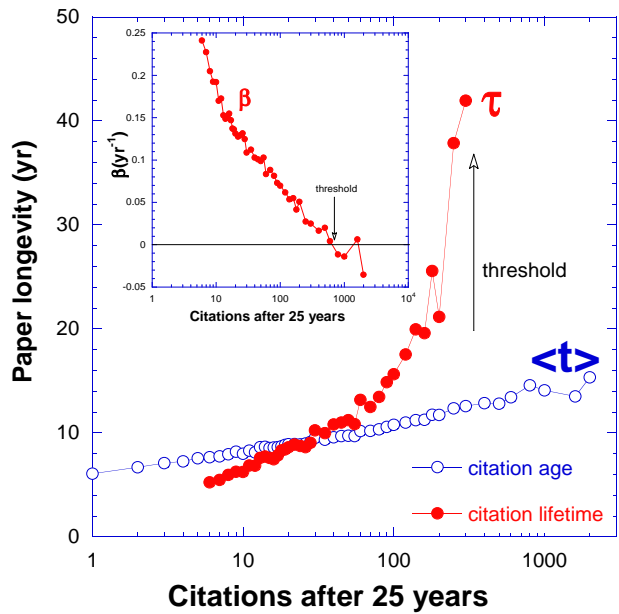}
\caption{Citation age $<t>$ (Eq. \ref{citation_age}) and  citation lifetime $\tau$ (Eq. \ref{age})  versus $k(t=25)$- the number of citations garnered by a paper after 25 years. The data were binned and each data point represents the average over one  bin. Note divergence of $\tau$ at the threshold of $k(t)\approx 600$. The inset  shows citation rate $\beta=1/\tau$ which changes its sign and becomes positive for $k(t)> 600$. The growth of $<t>$ and $\tau$ with $k$ is a signature of the \emph{nonlinear} autocatalytic process.
}
\label{fig:age}
\end{figure*}

Basing on these considerations we provide the experimental evidence for the nonlinear growth by analyzing various measures of the paper longevity.  One of them is a nonparametric measure  introduced by Redner \cite{Redner}, namely, "citation age" $<t>$. It is defined with respect to the paper age $t$ which is the number of years that passed after its publication. The citation age of the paper is nothing else but the mean age of the papers that cite it,
\begin{equation}
<t_{i}>=\frac{\int_{0}^{t} t_{i}dk_{i}}{k_{i}(t)}
\label{citation_age}
\end{equation}
In the extreme case when the citations  grow linearly in time,   $<t_{i}>=t/2$. Citation age exceeding  $t/2$  indicates accelerating growth while citation age below $t/2$ indicates some kind of saturation. Figure \ref{fig:age} shows citation age of the papers for $t=25$ years. We observe that  $<t>$ increases with the number of citations  as expected for the nonlinear growth mechanism, and eventually  achieves the critical value of $t/2=12.5$ years. This means that there is appreciable number of papers whose  citation dynamics  didn't come to saturation and they are actively cited even 25 years after publication!
%=25/2 exactly at the threshold of divergence of $\tau$, as expected. 
% In this latter case,  if the dynamics were exponential (Eq. \ref{age}), then in the long-time limit  $<t>\approx \tau+\Delta$. 

Another way to illustrate such exceptional behavior is to  approximate Eq. \ref{nonlinear}  by the exponential dependence 
\begin{equation}
k(t)=\frac{K}{\beta}\left[1-e^{-\beta(t-\Delta)}\right]
\label{age}
\end{equation}
where $K$ is some scale factor, $\Delta$ is a (small) delay  between the publication of the paper and the onset of citations, and $\beta$ is the rate. The latter  is negative when $k(t)$ accelerates with time and positive when $k(t)$ comes to saturation. In this latter case  $\tau=1/\beta$  has the meaning of citation lifetime. It is related to citation age (Eq. \ref{citation_age}) as follows: for the exponential dynamics  and in the long-time limit  $\tau+\Delta\approx <t>$. 

We measured $k(t)$ for all papers in our dataset, approximated it using Eq. \ref{age} and found microscopic parameters $\beta$ and $\Delta$.  Since these microscopic parameters strongly fluctuate, we binned all data into 40 logarithmically-spaced bins and considered the average  over the papers in each bin. The results are shown in Fig. \ref{fig:age}. The  rate  $\beta$ changes sign and becomes positive for highly-cited papers indicating accelerating citation dynamics. This change of sign occurs at the same threshold where citation age becomes equal to $t/2$ (Eq. \ref{citation_age}). The citation lifetime is $\tau\sim 5-6$ years for low-cited papers, for moderately- and highly-cited papers  citation lifetime increases and even diverges, as it is predicted by Eq. \ref{nonlinear}.

The divergence of the citation lifetime is a direct consequence of the nonlinear autocatalytic growth and it demonstrates  the tendency of citation network to develop a few hubs that attract the majority of citations. In the cumulative citation distribution these hubs appear  as "runaways" \cite{Golosovsky_EPL}.  These most highly-cited papers  have all chances to achieve infinite citation lifetime. This observation extends the well-known adage "the rich get richer"  to "the rich live longer".  
\section {Discussion}
\subsection{Why is the growth mechanism of citation networks so close to linear?}
When viewed from the perspective of  network dynamics, the preferential attachment mechanism does not favor  any particular value of the attachment exponent. From such perspective the ubiquity of linear or nearly linear preferential attachment is enigmatic. However, if we consider network dynamics from the perspective of a single node, the ubiquity of nearly linear preferential attachment appears naturally. 

Indeed the linear preferential attachment in the context of citations means that citation dynamics of the papers published in the same year has the same functional dependence and differs only in scale (we totally neglect here the stochastic character of the citation process). The difference between total number of citations of these papers is due to  initial conditions, namely  the number of citations that the papers garnered during first 2-3 years after publication. This is related to the number of readers which is determined by the journal's circulation. Since the majority of readers are graduate students who tend to copy once prepared reference list in all their publications, the citation lifetime  of a paper that some Ph.D. student came across, is the duration of his Ph. D. stay, namely 3-5 years.  Therefore, the initial impact of a paper on research  groups that undertook to cite it, usually continues for  1-2 generations of the Ph. D. students, namely for 6-10 years (see Fig. \ref{fig:age}). 

% The natural time scale in this context is the duration of the Ph. D. stay, namely 4-5 years.   That is why the vast majority of papers are cited during 10-15 years (the students continue to cite the paper in all their publications) while after 7-10 years the interest to this paper in each research group decays. and 7-10 years for the research group to remain adherent to one theme. 

If the above scenario were true for all papers, then the growth of the citation network would follow linear preferential attachment and the citation lifetime of all papers would be more or less the same. Figure \ref{fig:age} shows that while the citation lifetime of the vast majority of papers is indeed 6-10 years, there are quite a few papers that have much longer lifetime. We believe that these are the papers that induce "chain" reaction or cascade. Namely, the researchers can pick  up such paper not by reading the journal where it was published but through the impact of this paper on other research groups. In this case the paper starts it citation career in a new research group and its citation lifetime increases by another 6-10 years. Such process of spreading the ideas is similar to epidemiological process \cite {Goffman,Bettencourt} and to the redirection (copying) mechanism  \cite{copying,Simkin}. It seems that the papers whose impact propagates through the cascade process are responsible for the nonlinear growth of the citation network. The fraction of such papers in the whole pool of papers determines the degree of deviation of the attachment exponent from unity. The fact that this deviation is small, indicates that only a small fraction of papers ignites the chain reaction or cascade. Our measurements \cite{Golosovsky_PRL} indicate that these are the papers that garnered at least  $\sim 50-70$ citations at some moment in their citation career.

This cascade mechanism is specific for the citation network and it does not necessarily occurs in other networks.  Therefore, the growth mechanism of the complex networks other than citation network (Table I) can still follow linear  preferential attachment.

\section{Conclusions}
The dynamics of citation network is driven by the nonlinear autocatalytic growth with the attachment exponent $\alpha\sim1.2-1.3$.  The small but appreciable deviation of the growth process from linearity leads to a dramatic dynamical phase transition: the papers that exceed at some stage a certain number of citations become practically immortal: their citation lifetime diverges.  In the language of epidemiology these papers become endemic.

\section{Acknowledgments}
We are grateful to Filippo Radicchi, Alexander Petersen, Oleg Yordanov, and Andrea Scharnhorst for fruitful discussions.


\begin{thebibliography}{}
\bibitem{uncited} Consider a cohort of papers published in the same year. After a couple of years, when the general interest to this cohort already decayed, the annual number of citations gained by previously uncited papers is either 0 or 1. Therefore, the  mean annual number of additional citations gained by previously uncited papers is  $\overline{\Delta k_{i}}\approx\Delta N_{0}/N_{0}$ where $N_{0}$ is the number of uncited papers and $\Delta N_{0}=N_{0}(t)-N_{0}(t+1)$ is the number of uncited papers that got their first citation during recent year $t+1$. If the total number of papers in the dataset is $N$, then $\overline{\Delta k_{i}}\approx\Delta P_{0}/P_{0}$ where $P_{0}=N_{0}/N$ is the fraction of uncited papers.
\end{thebibliography}
\end{document}